\documentclass[usenatbib,useAMS]{mn2e}
\usepackage{times}
\usepackage{epsfig}

\usepackage[usenames,dvips]{color}

\begin{document}
\title[Simulation of a strategy for the pixel lensing of M87 using the Hubble Space Telescope]
{Simulation of a strategy for the pixel lensing of M87 using the
Hubble Space Telescope}

\author [Sedighe Sajadian and Sohrab Rahvar]
{Sedighe Sajadian$^{1}$ and Sohrab Rahvar$^{1,2}$ \thanks{rahvar@sharif.edu} \\
$^1$ Department of Physics, Sharif University of Technology, P.O.Box
11365--9161, Tehran, Iran \\
$^2$ Perimeter Institute for Theoretical Physics, 31 Caroline Street
North, Waterloo, Ontario N2L 2Y5, Canada}

\maketitle

\begin{abstract}
In this work we propose a new strategy for the pixel-lensing
observation of M87 in the Virgo cluster using the {\it Hubble Space
Telescope} (HST). In contrast to the previous observational strategy
by Baltz et al., we show that a few days intensive observation with
the duration of $\sim 90~$min in each HST orbit will increase the
observational efficiency of the high-magnification events by more
than one order of magnitude. We perform a Monte-Carlo simulation for
this strategy and we show that the number of high magnification
microlensing events will increase at the rate of $4.2$ event per day
with a typical transit time scale of $16$ h. We also examine the
possibility of observing mini-halo dark matter structures using
pixel lensing.

\end{abstract}

\section{introduction}
One of the consequence of Einstein General Relativity is that light
is deflected by the gravitational field of an object with an angle
of
\begin{equation}
\alpha= \frac{4GM}{c^{2}b},
\end{equation}
where $M$ is the mass of the deflector and $b$ is the impact
parameter of the light ray \cite{ei11}. In the case of lensing of a
background star by another star, this deflection creates extra
images of a background object where the images are too close to be
resolved by the ground-base telescopes. This type of lensing is
called gravitational microlensing. In 1936, Einstein said "there is
a little chance of observing gravitational lensing caused by the
stellar mass lenses". However, after several decades, using advanced
cameras and computers, and following the propose of Paczy\'nski
(1986), the first microlensing event has been discovered by Alcock
et al. (1993).

The aim of the microlensing observational groups such as MACHO, EROS
and OGLE \cite{al00,af03,wyr09} has been to monitor stars the Large
and the Small Magellanic Clouds in order to detect the massive
compact halo objects (MACHOs) of the halo using gravitational
microlensing. The passage of the MACHOs from the line of sight of
the background stars causes an achromatic magnification of the
background stars with a specific shape. Comparing the number of the
observed events with that expected from the halo models have shown
that MACHOs made up less than 20 per cent of the Galactic halo
\cite{mil01,rah04,mon09}. They can not cover all the dark matter
budget of the halo. In this paper, we question further the existence
of MACHOs on cosmological scales, and suggest a strategy for
observation using the {\it Hubble Space Telescope (HST)}.

Gravitational microlensing towards the source stars beyond the
Galactic halo can be performed with an observational method -
so-called pixel lensing- which is different to the conventional
method. When the background stars are too far and the field is too
dense to be resolved, using this method a high magnification
microlensing event can change the light flux in each pixel. By
recording the time variation of the light in each pixel, it is
possible to recognize a microlensing event \cite{cr92,ba93}. An
important factor for increasing the sensitivity of the observations
when using the pixel lensing is the size of point spread function
(PSF). Several groups as MEGA, AGAPE and ANGSTROM, have monitored
M31. They have found few pixel-lensing candidates in this direction
\cite{cr96,an99,ke06}.

Gould (1995) proposed the extension of pixel lensing to cosmological
scales, using the observation by the HST towards the Virgo cluster.
He proposed an investigation of the intercluster MACHOs and obtained
a theoretical optical depth of $\tau\sim 3\times 10^{-5} f$,  where
$f$ is the fraction of halo composed of MACHOs \cite{gou95}. The
detection threshold of an event was defined by the accumulation of
the signal to noise of the events for the duration that an event is
observable. So, in this observation, one would expect to detect
microlensing events with a long duration and a low peak in light
curve. The estimation obtained for the rate of events with this
observational threshold was $\Gamma \sim 18 f~day^{-1}$. Baltz et al
(2004) used the {\it HST} for pixel lensing and monitored M87 for
one month, using with the strategy of taking one date point per day.
After data reduction, seven candidates remained. With further
investigation of the light curves in two different colors they
concluded that one candidate could be a microlensing event.
Comparing the observational result with the expectation from
generating synthetic light curves, they concluded that the fraction
of MACHOs in both the Milky Way and the Virgo halos are almost equal
\cite{al00}. However, the problem with this statistical analysis is
that one microlensing event has a large statistical uncertainty. In
order to have a better estimation of the MACHOs in the Virgo
cluster, we need to be observed more events.

In this work we extend the work by Gould (1995) and Baltz et al.
(2004), and we propose an alternative observational strategy using
the {\it HST}. Our suggestion is to observe the very high
magnification events in M87, which has a variation in its light
curve in the order $1$ d. In order to improve the sensitivity of
observing of short duration events, we need to increase the sampling
rate by the order of $1$ h. We propose intensive short-duration
observation of M87 of the order of $1$ h using the Wide Field
Planetary Camera 3 (WFC3) of the {\it HST}. We suggest taking one
date point in each orbit and collecting about $15$ photometric data
point per every $24$ h. This strategy will enable us to detect very
high magnification short-transit pixel lensing events.

First, we make a rough calculation to estimate the optical depth and
the detection rate of the events using this observational strategy.
In order to have a better estimation, we continue with a Monte-Carlo
simulation for three observational programs of $1$, $2$ and $3$ d.
We examine how sensitivity the results are to the duration of the
observational program. We show that with $1$, $2$ and $3$ d of
intensive observation with the {\it HST}, it is possible to detect
$4.2$, $13.4$ and $18.5$ events per day, respectively. We also
simulate the observation using the strategy of Baltz et al. (2004),
with the cadence of $1$ d, and find $0.02$ event per day for one
month observations.

The paper is organized as follows. In section \ref{theopt}, we
introduce the Virgo cluster and obtain the theoretical optical depth
and the event rate of the high magnification events from structures
along the line of sight. In section \ref{mont}, we perform a
Monte-Carlo simulation to obtain the number of the high-
magnification events found with the {\it HST}. The results are given
in section \ref{results}. In section \ref{conc} we give our
conclusion.

\section{pixel lensing of M87}
\label{theopt}

Virgo cluster contains $\sim 1300$ galaxies and is located about
$16.5$ Mpc from us. It has am apparent size of $8^{\circ}$ and
coordinate of RA$\sim 12~hrs$ and DE$\sim 12^{\circ}$ \cite{fo01,
me07}. The brightest galaxy of this cluster is M87 which is located
at the center of this structure. Similar to the Local Group, the
mass of the Virgo cluster is mainly made up of dark matter. With
microlensing of the Virgo cluster, it could be possible estimate the
fraction of MACHOs in the halo of this structure.

Because stars in the Virgo cluster cannot be resolved using present
ground- and space- based telescopes, the standard microlensing
technic cannot be used for the stars in the Virgo cluster. To make
an estimation from the column density of the stars, we assume that
galaxies in the Virgo cluster have the same surface density of stars
as in the Milky way, ($\Sigma = 50 M_{\odot}/pc^2$). For {\it HST}
with a PSF size of $0.067$ arcsec in the WFC3, the number of stars
inside the PSF would be about $1000$ \cite{wfc3}. To detect a
microlensing event in a high blended background, we need a very high
magnification event to change the flux of pixels inside the PSF. In
what following, we provide a rough estimation from the optical depth
and the event rate in terms of the transit time towards M87. In
Section 3, we obtain a precise value of the event rate, based on a
Monte Carlo simulation using {\it HST} observations.

The standard definition of the optical depth is the cumulative
fraction of area covered by the Einstein ring for all the distances
from the observer to the source star \cite{pa86}:
\begin{equation}
\label{to} \tau = \int_{0}^{D_{s}} \frac{\rho(D_{l})}{M} \pi
R_{E}^{2} dD_{l},
\end{equation}
where $R_{E}$ is Einstein radius, $M$ is the typical mass of the
lens, $\rho(D_{l})$ is the mass density of the lens and $D_{l}$ is
the distance between the observer and lens.

\begin{figure}
\begin{center}
\psfig{file=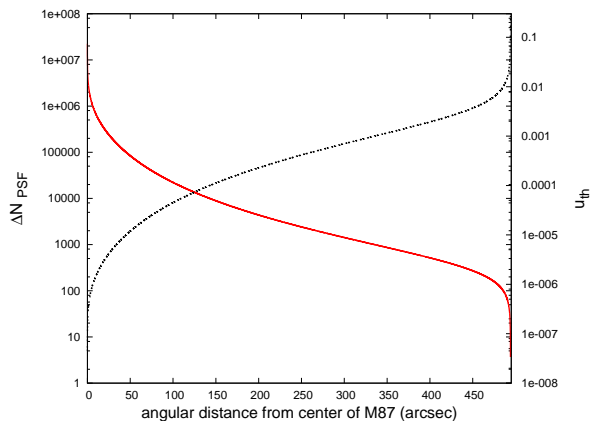,angle=0,width=8.cm,clip=} \caption{Number of
stars inside the PSF size (solid line) indicated at the left hand
side of y-axis and threshold for the impact parameter (dashed-line)
indicated at the right hand side as a function of angular distance
from the center of the galaxy.} \label{npsf}
\end{center}
\end{figure}

Taking M87 as the target galaxy, the column density of stars at the
center of this galaxy is much larger than that at the edge of the
galaxy. Thus, the threshold for detecting a very high magnification
event would depend on the position of the source star in M87.
Assuming a spherical model for M87, we calculate the column density
of the stars by integrating the number density of stars along the
line of sight, $\Sigma(r) =\int n(r,z) dz $. Then, the number
density of stars for a given solid angle would be $dN/d\Omega = D^2
\Sigma$. The number of stars inside the PSF is given by $\Delta
N_{PSF} = dN/d\Omega\times \Omega_{PSF}$. Fig. (\ref{npsf}) plots
the number of stars inside the PSF in terms of separation from the
center of the galaxy in arcsec, using the galaxy model in McLaughlin
(1999).
The threshold impact parameter can be defined according to the
blending inside the PSF. Because the threshold magnification should
be more than the number of stars inside the PSF (i.e. $A_{th}>\Delta
N_{PSF}$), the corresponding threshold for the impact parameter can
be defined as $u_{th}<(\Delta N_{PSF})^{-1}$. Fig. \ref{npsf} shows
the threshold impact parameter as a function of the angular
separation from the center of the galaxy, as indicated by the dashed
line.

We can define a new optical depth according to the threshold impact
parameter in such a way that $\tilde{R}_E = u_{th} R_E$. Using the
definition of the optical depth based on this reduced Einstein
radius, this relates to the conventional definition of optical depth
as follows:
\begin{equation} \tilde{\tau}(\hat{n}) =
u_{th}^{2}(\hat{n}) \tau(\hat{n}), \label{obsopt}
\end{equation}
where, $\hat{n}$ is the direction of the observation and $\tau$ is
the standard definition of the optical in equation (\ref{to}). In
what follows, we estimate the numerical value of the standard
optical depth, resulting from the various structures along the line
of sight. Finally, we obtain the optical depth of the observable
high-magnification events by multiplying the optical depth by the
threshold impact parameter, as in equation (\ref{obsopt}). We note
that the lenses along the line of sight can be in the halo of Milky
Way, the disc of the Milky Way, the halo of the Virgo cluster, the
halo of M87 or M87 itself. We calculate the optical depth towards
all area of M87 and we provide the average optical depth of each
structure, as follows.

(i) We assume lenses in the galactic disc as the nearest structure
to the observer. The mass density for the lenses is given by
\cite{bin}
\begin{equation}
\label{dd} \rho(R,z) =
\frac{\Sigma}{2H}\exp\left[-\frac{(R-R_{\odot})}{h}\right]
\exp\left(-\frac{|z|}{H}\right).
\end{equation}
Here, $h$ is the length-scale of the disc, H is the height-scale and
$\Sigma$ is the column density of the disc at the position of the
sun, $R_{\odot}$. Substituting $\rho (R,z)$ into the equation
(\ref{to}) and integrating over $D_{l}$, the average value of the
optical depth weighted by the number of background stars over M87,
we obtain $\tau_{i} \sim 1.4 \times 10^{-9}$.

(ii) For the Galactic halo, we use NFW mass density profile
\cite{nfw}
\begin{eqnarray}
\label{rr} \rho(r)=f_{h} \frac{\delta_{c}
\rho_{c}}{(r/r_{s})(1+r/r_{s})^{2}},
\end{eqnarray}
Here, $r_{s}$ is the characteristic radius, $\rho_{c}={3H_0^{2}}/{8
\pi G}$ is the present critical density, $f_{h}$ is the fraction of
halo composed of MACHOs, $r$ is the position of the lens from the
center of galaxy and $\delta_{c}=200c^{3}g(c)/3$, where $c$ is the
concentration parameter and $$g(c)=\frac{1}{\ln(1+c)-c/(1+c)}.$$ We
set $c=18$ and $r_{s}=14 kpc$ for galactic halo \cite{ba05}. We
obtain the average optical depth from this structure $\tau_{ii} \sim
f_h \times 1.1\times 10^{-6}$, which is three order of magnitude
larger than that of the Galactic disc.

(iii) The third structure is the halo of Virgo cluster. We set the
NFW parameters for the Virgo halo as $r_{s} \sim 560 kpc$ and
$\delta_{c}\rho_{c}=3.2\times 10^{-4} M_\odot/pc^3$\cite{mc99}. The
optical depth is obtained as $\tau_{iii}\sim f_v\times 3.38\times
10^{-5}$, where $f_v$ is the fraction of Virgo halo composed of
MACHOs. We note that the optical depth from the Virgo cluster is
also one order of magnitude larger than the Galactic halo.
\begin{figure}
\psfig{file=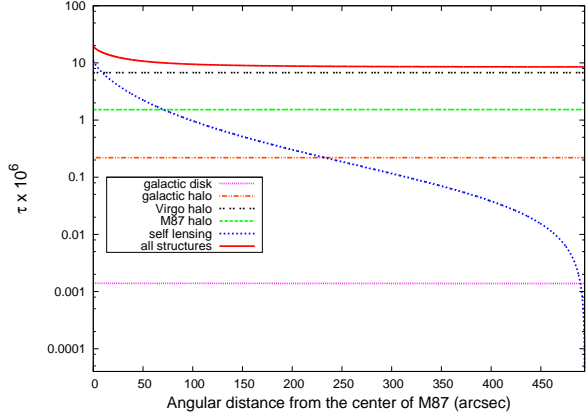,angle=0,width=8.cm,clip=} \caption{The optical
depth towards the Virgo cluster in terms of the transverse angular
separation from the center of the cluster for various structures.
The solid line is the overall optical depth by accumulating the
result from the different structures. } \label{optdepth}
\end{figure}

(iv) The lens can reside in the galactic halo of M87. We take the
NFW model for the halo of this structure with the parameters of
$r_s= 50 kpc$ and $c= 15$ \cite{ge09,do09}. The optical depth
resulting from this structure is $\tau_{iv} \sim f_{M87}
\times7.62\times 10^{-6}$.

(v) Finally, the lens can be in M87, so-called self-lensing events.
We assume M87 is located at the center of the Virgo cluster with a
spherical shape and a radius size of $40 kpc$ \cite{ge09}. For this
structure, we use the mass density profile of \cite{mc99}:
\begin{equation}
\label{mc}\rho(r) =\frac{3-\gamma}{4}\frac{\Upsilon_{B}L_{B}}{\pi
a^3}\left(\frac{r}{a}\right)^{-\gamma}\left(1+\frac{r}{a}\right)^{(\gamma-4)},
\end{equation}
where $\gamma=1.33$, $a=5.1\pm0.6~kpc$, $L_{B}=(5.55\pm0.5)\times
10^{10}~L_{\odot,B}$, $\Upsilon_{B}=14.6\pm0.2
~M_{\odot}L_{\odot,B}^{-1}$ and $r$ is the distance from the center
of M87. The optical depth resulting from the self-lensing is
$\tau_{v}\sim 1.73 \times 10^{-6}$.

\begin{figure}
\psfig{file=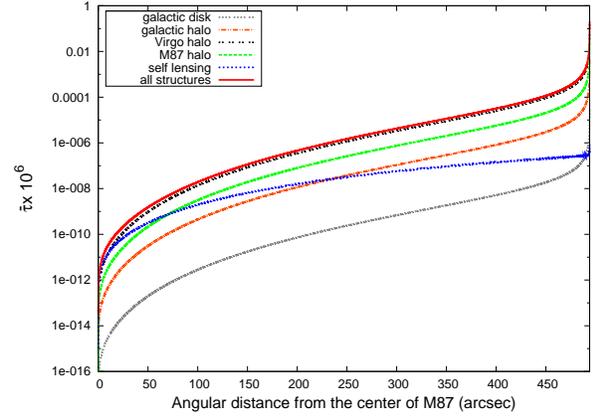,angle=0,width=8.cm,clip=} \caption{ The reduced
optical depth as a function of angular distance from the center of
M87, resulting from the multiplication of the optical depth to the
square of threshold impact parameter, i.e. $\tilde\tau(\theta) =
\tau(\theta)u_{th}^2(\theta)$.} \label{obsod}
\end{figure}

We note that the halo of Virgo has the highest contribution in the
optical depth towards M87. Fig. (\ref{optdepth}) shows the optical
depth resulting from the various structures as a function of angular
separation from the center of the galaxy. Knowing the optical depth
and the threshold impact parameter, we use the definition of the
reduced optical from equation (\ref{obsopt}) and we obtain
$\tilde\tau$ in terms of angular distance from the center of M87 by
multiplying the standard definition of the optical depth in
Fig.(\ref{optdepth}) by $u_{th}^2(\theta)$ from Fig. \ref{npsf}. The
result is shown in Fig.\ref{obsod}. We summarize the result of the
reduced optical depth in Table 1. Here, we assume that MACHOs
contribute $20$ per cent of the halo mass.

\begin{figure}
\psfig{file=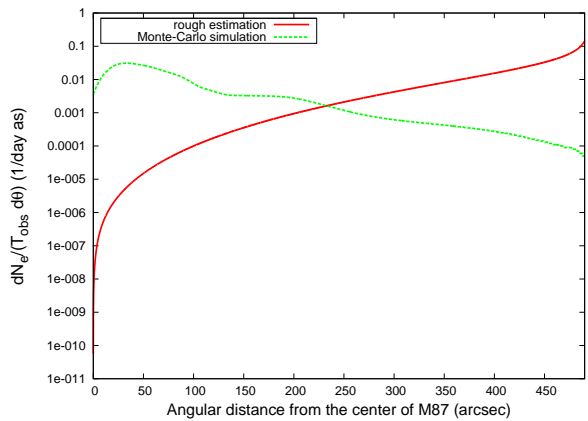,angle=0,width=8.cm,clip=} \caption{ The event
rate as a function of angular distance from the center of M87. The
solid line is obtained from the rough calculation and the dashed
line shows the result from the Monte-Carlo simulation.}
\label{event}
\end{figure}

We note that, in contrast to the standard definition of the optical
depth, which is almost constant for all the directions towards the
target structure, the reduced optical depth depends on the line of
sight towards M87. It is small at the center of M87 and becomes
larger at the edge of the structure.
\begin{table*}
\begin{center}
\begin{tabular}{|c|c|c|c|c|c|c|c|}
 & & a  & b & c & d & e & overall \\
\hline\hline
&$\tilde{\tau}\times 10^{11}$& 0.00074  & 0.118 & 3.622 & 0.819
& 0.0028 & 4.563 \\
\hline

&$t_{1/2}(hr) $& 6.11 & 5.39 & 17.32 & 9.02 & 32.26 &15.85\\

\hline
&$N_{e}(1/day)$& 0.0032 & 0.607 & 5.794 & 2.513 & 0.013 & 8.93\\

\label{tab1}
\end{tabular}
\label{table1}
\end{center}
\caption{Reduced optical depth (first row), transit time (second
row) and the number of the events (third row) for each structure.
Structures in each column are noted by (a) Galactic disk; (b)
Galactic halo; (c) Virgo halo; (d) M87 halo; (e) self lensing and
last column is mean value for the optical depth and transit time and
the the overall number of detected events per day.}
\end{table*}
From the definition of the optical depth, we can obtain the rate for
events for those with the impact factor smaller than $u_{th}$. From
the dependence of the number of events on the optical depth, the
rate of events is obtained as
\begin{equation}
\frac{N_e}{T_{obs}N_{bg}}=\frac{2}{\pi}
\frac{\tilde{\tau}}{\tilde{t}_{E}} , \label{tau}
\end{equation}
Here, $N_e$ is the number of the observed events, $N_{bg}$ is the
number of the background stars during the observational time of
$T_{obs}$ and the reduced Einstein crossing time is defined as
$\tilde{t_E} = u_{th} t_E$. Using the definition of $\tilde{\tau}$,
the rate of events can be written in terms of the optical depth and
the Einstein crossing time as
\begin{equation}
\frac{N_e}{T_{obs}N_{bg}}=\frac{2}{\pi} \frac{\tau}{t_{E}} u_{th}.
\label{tau2}
\end{equation}
One of the relevant time scales in the high-magnification
microlensing events is ${t}_{1/2}$, which is defined as the full
width at half-maximum (FWHM) timescale of the high magnification
events. For high magnification events, the magnification changes
with the impact parameter as $A = 1/u$. The transit time for these
events relates to the Einstein crossing time as  $t_{1/2}= 2\sqrt{3}
u_0 t_E$. Hence, for each event, after fitting the observational
light curve with the theoretical light curve, we can obtain
$t_{1/2}$ in terms of $u_0$ and $t_E$. Finally, taking the number
density of background stars $dN(\theta)_{bg}/d\Omega$ as a function
of angular distance from the center of the structure, we can obtain
the differential rate of events as
\begin{equation}
\frac{ dN_e}{T_{obs} d\theta }= \frac{4 dN_{bg}(\theta)}
{d\Omega}\frac{u_{th}(\theta)\tau(\theta)}{t_E(\theta)}\theta.
\label{dfr}
\end{equation}
Using this equation and the estimation for the time-scale of $t_E$,
we obtain the rate of events as shown by the solid line in
Fig.(\ref{event}) in terms of the angular distance from the center
of M87. Integrating over $\theta$ results in the number of the
detectable events. Here, we obtain about $\sim 9$ high magnification
microlensing events per day. We note that in this rough estimation,
we did not use the quality of light curve on delectability of the
events. Also we ignored the finite-size effect on the detections
made with pixel lensing. In the section 3, we will re-analyse and
simulate the light curves of M87 using the {\it HST}. Table (1)
reports the reduced result of this preliminary analysis and provide
the reduced optical depth from the various structures, $t_{1/2}$ and
and the rate of the events.

\begin{figure}
\psfig{file=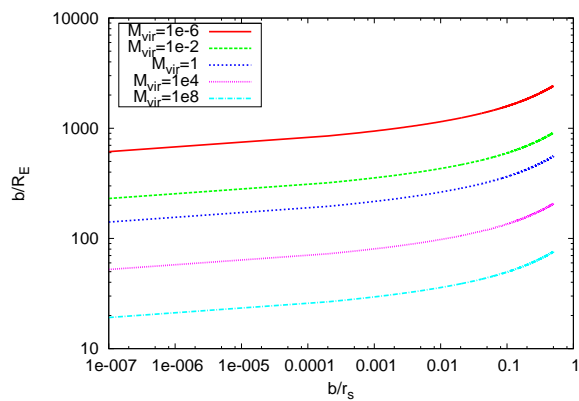,angle=0,width=8.cm,clip=} \caption{ The
impact parameter normalized to the Einstein radius for micro-halos
in terms of the impact parameter normalized to the size of
micro-halo for the structures with various masses.}\label{halo}
\end{figure}

Finally we examine the possibility of observing dark matter
micro-halo structures \cite{goe07,eric01} using pixel lensing.
Taking the NFW profile for these structures, in order to have high-
magnification events from these lenses, we have to decrease the
impact parameter down to $\sim 10^{-3}$. The problem with dark halo
structures is that the mass of the structure enveloped by sphere
with the radius of the impact parameter decreases towards the center
of structure. Fig. (\ref{halo}) plots the impact parameter
normalized to corresponding Einstein radius in term of the impact
parameter normalized to the size of the structure. For microhalos,
the normalized impact parameter, $b/R_E$ is always larger than one
and never reach below $10^{-3}$, a favorable value for the pixel
lensing. Our conclusion is that we can ignore possible contribution
of these structures in pixel lensing. With current technology, it is
not possible to detected the micro-halo structures.


\section{MONTE-CARLO SIMULATION}
\label{mont} In this section we perform a Monte-Carlo simulation to
obtain the detection efficiency and the rate of microlensing events
towards M87 using pixel lensing. In the first step, we simulate the
stellar distribution in M87, using the color-magnitude profile of
the stars of the galaxy. Then, we use the foreground structures to
obtain the distribution of the lenses along the line of sight. In
the pixel lensing, the magnification is suppressed by the blending
and the finite size effect. We take into account these two effects
in our simulation.

\begin{figure}
\psfig{file=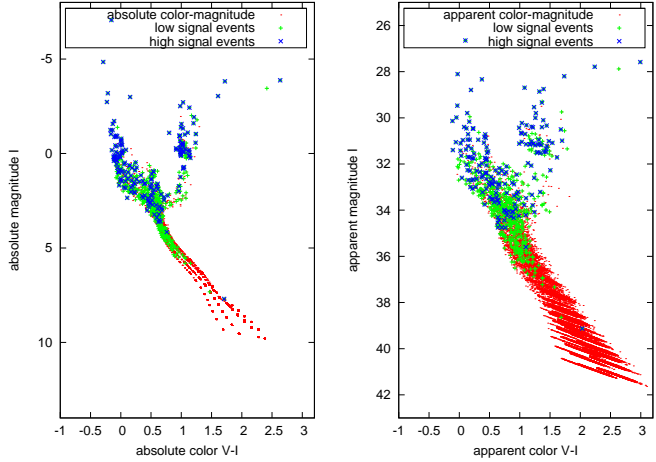,angle=0,width=9.cm,clip=} \caption{CM diagram
generated by Padova Isochrones. The left panel is the absolute
color--magnitude diagram and the right panel is the apparent CM
generated according to the distance modulus, extinction and
reddening of the stars. We show a sample of microlensed stars with a
high signal to noise of $Q=30$ (cross sign) and low signal to noise
of (plus sign) $Q=5$ in the Pixel lensing.} \label{cm}
\end{figure}

To simulating the stellar distribution in M87, we use the
theoretical Isochrones of the color-magnitude diagram diagram (CMD)
obtained using the numerical simulation of the stellar evolution
from the Padova model \cite{padova}. The Isochrones are defined for
the stars with different mass, color, magnitude but with the same
matalicity and age. To simulating the CMD, we select isochrones with
the metalicity range of $Z= 0.0004, 0.008, 0.004, 0.001, 0.019$ and
$0.030$. For the evolution of the stars, we take the ages of the
stars in the range of $log(t/yr)= [6.6,10.2]$ with interval of
$\bigtriangleup log(t)=0.05$~dex. For each age range, we take a
metalicity according to the age-metalicity relation, and the
distribution function for the metalicity from Twarog (1980). In
order to generate the CMD of the stars at the present time, we need
an initial mass function of the stars and the star formation rate
function for M87. The initial mass function is chosen from the
Salpeter function \cite{sa95}. For the star formation rate, we
assumed a positive rate $dN/dt>0$ similar to that in our Galaxy
\cite{cig,karami}. The result of our simulation in the absolute CMD
is shown in the left panel of Fig.(\ref{cm}).

In order to generate the apparent color-magnitude of the stars, we
calculate the distance modulus of each star in M87 and add it to the
corresponding extinction. Because dust in the interstellar medium is
proportional to the Hydrogen density, we can use the column density
of hydrogen $(N_H)$ along the line of sight as an indicator of
amount of extinction. For the $V$ and $I$ bands, the amount of
extinction is given by \cite{we01}:
\begin{eqnarray}
\label{tt} A_{V}(mag)&=&5.3\times10^{-22} N_{H}(atom/cm^{2}), \nonumber \\
A_{I}(mag) &=& 2.6\times10^{-22} N_{H}(atom/cm^{2}).
\end{eqnarray}
For the distribution of the hydrogen in M87, we choose the
theoretical distribution function of hydrogen, which has the best
fit to the X-ray observation \cite{ts94}:
\begin{equation}
\label{ss}n_{e}(r)= n_{0}\frac{(r/a_{1})^{-\alpha}}{1+(r/a_{1})}.
\end{equation}
Here, $a_{1}=9.51 kpc$, $\alpha=0.36$, $n_{0}=3.51\times10^{-2}
atom/cm^{3}$ and $r$ is the distance from the center of the
structure.

Finally, the apparent magnitude and color of each star are given by
\begin{eqnarray}
m_{V} &=& M_{V}+5\log\left[\frac{d(pc)}{10}\right] +A_{V}, \nonumber\\
C_{obs} &=& C_{0}+ A_{V}-A_{I},
\end{eqnarray}
Here, $C_{0}$ is the absolute color, $M_{V}$ is the absolute
magnitude in the visible band and $d$ is distance of stars in
parsec. We plot the apparent color-magnitude of the stars of M87 in
the right panel of Fig.(\ref{cm}), including the effect of the
distance modulus, extinction and the reddening.

\begin{table*}
\begin{center}
\begin{tabular}{|c|c|c|c|c|c|c|c|c|c|}
& & & a  & b & c & d  & e & overall\\
\hline \hline

&1&$\tau\times 10^{6}$& 0.00139  & 0.219 & 6.758 & 1.525 & 1.734 & 10.238\\
\hline

&2&$N_{e} (1/day)$& 1.2e-6 & 0.0022 & 4.2 & 0.018 & 0.0072 & 4.2 \\

\hline\hline
& & & a  & b & c & d  & e & average\\
\hline\hline

&3&$\rho_{\star}/u_{0}$ & 0.002  & 0.16 & 1.5 & 6.2 &  1.3 & 1.5 \\
\hline

&4&$t_{1/2} (hr)$& 12.6 & 15.6 & 16.9 & 5.6 & 4.5 & 16.3 \\
\hline

&5&$A_{max} $& 447.9 & 688.9 & 319.8 & 509.7 & 537.9 & 348.4 \\
\hline

&6&$u_{0}$& 0.004 & 0.007 & 0.032 & 0.003 & 0.0038 & 0.029 \\
\hline

&7&$m (mag)$& 31.3 & 30.4 & 29.6 & 31.1 & 32.2 & 29.7 \\
\hline

\end{tabular}
\label{tab2}
\end{center}
\caption{Result from the simulation of one day observation by the
HST with the WFC3 camera. The columns are (a) Galactic disk;(b)
Galactic halo; (c) Virgo halo; (d) M87 halo; (e) self lensing. The
first and second row represents the optical depth and the number of
the events resulting from each structure and the last column
indicates the overall numbers. The third line is the mean value of
$\rho_{\star}/u_{0}$ for the observed events in each structure. The
forth row represents the average transit time scale of the events,
the fifth row is the average of maximum magnification, the sixth row
is the impact parameter and the seventh row is the apparent
magnitude of the source stars. The last column represents the
average value for the parameters of the lenses.
 \label{tab2} }
\end{table*}

In the next step, we generate the light curve of the microlensing
events by choosing the parameters of the lens as the mass, the
velocity and the location, according to the corresponding
distribution functions. The location of the lenses from the observer
is calculated according to the probability function for the location
of lenses
$$d\Gamma/dx \propto \rho(x)\sqrt{x(1-x)}$$,
where $x=D_{l}/D_{s}$ and $\rho(x)$ is the mass density profile of
the lenses along the line of sight. The mass of the lenses, in units
of solar mass, are chosen from the Kroupa mass function
\cite{kroupa1, kroupa2}, $\xi \propto M_{l}^{-\alpha}$, where
$\alpha= 0.3$ for $0.01\leq M_{l}<0.08$, $\alpha= 1.3$ for $0.08\leq
M_{l}<0.5$ and $\alpha= 2.35$ for $0.5\leq M_{l}\leq 1.0$. The
velocity of lenses in the halo is chosen from the Maxwell-Boltzmann
distribution with the dispersion velocity indicated by the Virial
theorem. For the Galactic halo, the Virgo halo and the M87 halo, the
dispersion velocities are set as $\sigma \sim 156~kms^{-1},
1000~kms^{-1}$ and $350~kms^{-1}$, respectively \cite{de06, ge09}.
The distribution for velocity of stars in the Galactic disc is also
chosen from ellipsoid of the dispersion velocity \cite{bin}. The
velocity of stars in M87 is chosen from Maxwell-Boltzmann function
with $\sigma \sim~360 kms^{-1}$ \cite{ro01}. The minimum impact
parameter $u_0$ in this simulation has to be chosen geometrically
uniform in the range of $u_{0} \in [0,1]$. However, in the
Monte-Carlo simulation, we generate a uniform impact factor in the
logarithmic scale to have enough statistics of the high-
magnification events and to reduce the computation time. In order to
recover the uniform distribution of impact parameter, we consider a
weight proportional to the impact parameter $u_0$ when counting the
number of events.


To calculate the finite-size effect, we need the size of the stars
projected on the lens plane. For the main-sequence source stars, the
radius of the stars is obtained according to the mass--radius
relation as
\begin{equation}
R_{\star} = M_{\star}^{0.8}.
\end{equation}
Here, the parameters are normalized to the Sun's values. Red giants
(RGs) also follow the relation of \cite{haya62}
\begin{equation}
M_{RG}^{1/2}R_{RG}^{3/2}=const.
\end{equation}
The formalism of the finite-size effect in the gravitational lensing
was obtained by Maeder \cite{mae73} and developed later for
microlensing by Witt \& Mao \cite{wit}. The relevant parameter in
the finite size effect is $\rho_{\star}/u_{0}$, where $\rho_{\star}$
is the projected radius of the star on the lens plane normalized to
the Einstein radius. For the population of the source stars in M87
and lenses in various structures along the line of sight, Table
(\ref{tab2}) gives the average value of $\rho_{\star}/u_{0}$, with
the corresponding $\rho_{\star}$ and $u_{0}$ for the observed events
in each structure. We examine the finite-size effect on the main
sequence and RG stars. We find that $43\%$ per cent of stars with an
absolute magnitude less than zero are RG, but for the microlensed
stars in this range only $17$ per cent are RGs.


In the pixel lensing, because of the high blending in each pixel,
the signal to noise ratio from microlensing must be so high to be
distinguishable from the background brightness of the host galaxy.
For a background star with the apparent magnitude of $m$, the extra
photons during the magnification of the star are obtained by
\begin{eqnarray}
\label{sig} \delta N = (A-1) \times t_{exp} \times
10^{-0.4(m-m^{zp})}.
\end{eqnarray}
Here, $A$ is magnification factor of the source star, $t_{exp}$ is
exposure time and $m^{zp}$ is the zero-point magnitude equivalent to
the flux $= 1$ photoelectron$^{-1}$, which is equal to $25.10$ and
$26.08$ mag in {\it I}-band (F814W) and {\it V}-band (F606W),
respectively, for the WFC3 camera of the {\it HST}. We note that the
observation by Baltz et al. (2004)
were performed with the WFC2, which had a lower zero-point magnitude
of $23.86$ mag for the {\it I} band. The exposure time can be
obtained from the visibility of M87 by HST. For M87, which is
located at $DEC = 12^\circ$, the visibility would be about $52~$min
in each {\it HST} orbit. We use the {\it I}-band (F814W) filter for
observation in the simulation. We show that the background events,
such as novas, cannot pass our light curve criterion for a short
observation period. Thus, we do not use another filter, such as the
{\it R} band, for the variable identifer and devote all the
observation time in the {\it I} band to increase the signal-to-noise
ratio. Within each orbit, we assume six $260~$s exposure with an
overall exposure of $1560~$s in {\it I} band. The noise is
considered as the intrinsic Poisson fluctuation inside the PSF. This
is proportional to the square of the number of the photons received
from the background sky and the host galaxy:
\begin{equation}
\label{sss} \sqrt{N} = 10^{0.2 m^{zp} }
\sqrt{\Omega_{PSF}(10^{-0.4 \mu}+10^{-0.4\mu_{sky} }) t_{exp} },
\end{equation}
Here, $\Omega_{PSF}$ is the characteristic area of the PSF,
$\mu_{sky}$ is the brightness of the background sky (set to $21.6
mag~as^{-2}$ and $\mu$ is the surface brightness profile of M87. In
our Monte-Carlo simulation, we accept the events with the following
criterion:
\begin{equation}
\label{qq} Q=\sqrt{\sum_i (\delta N_i/\sqrt{N})^{2} }> Q_{crit}.
\end{equation}
Here, the summation is performed over the observed points of the
light curve \cite{go96} and $Q_{crit}$ is the critical value for the
signal-to-noise of the detection (chosen to be $Q_r = 5$ for a
relaxed criterion and $Q_s=30$ for a strict criterion). The $Q^2$ is
equivalent to the $\chi^2_{bl}$ fit of the light curve to the
baseline.

As the second criterion, we adopt those events in which the peak of
the light curve resides within the duration of the {\it HST}
observation, (i.e. $t_0\in[0,T_{obs}]$, means that we detect events
with both rising and declining parts of the light curve). For the
short term events, the probability of fulfilling this criterion is
higher than for the long duration events.
We add another criterion, which is sensitive to the fast variation
of the light curve, for which we detect at least three consecutive
data points, $3\sigma$ above the background flux. This means that we
have, at least, a $3\times 3\sigma$ deviation of the microlensing
light curve from the bassline fit. In other words, the difference
between the two different fits should be at least $\Delta\chi^2 =
\chi^2_{bl} - \chi^2_{ml} > 27$.

\begin{figure}
\psfig{file=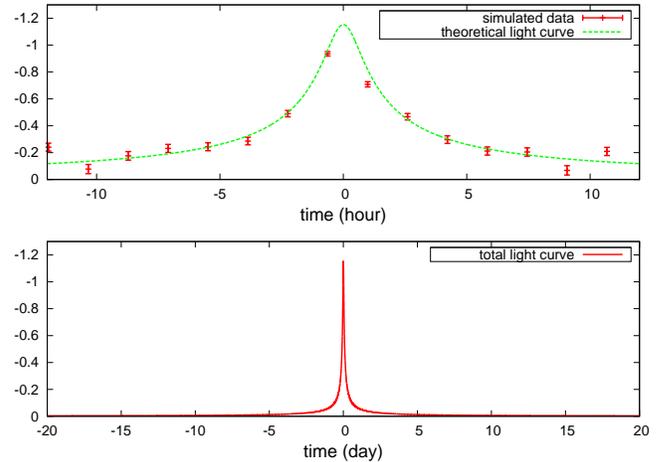,angle=0,width=9.cm,clip=} \caption{The theoretical
light curve of a high magnification microlensing event including the
blending of the background stars with the simulation of the
observation with the cadence of $\sim 90~min$ (top panel data
points). In the bottom panel we show the same event for the duration
of $40$ days.}\label{light}
\end{figure}

\begin{figure}
\psfig{file=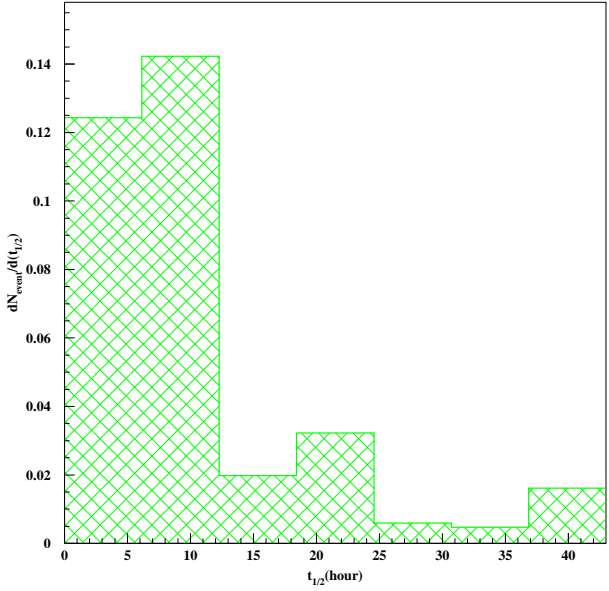,angle=0,width=9.cm,clip=} \caption{ The
distribution function of $t_{1/2}$ of the observed events with the
criterion of $Q>5$. \label{time}}
\end{figure}

In Fig.(\ref{cm}), we identify the observed events that result from
the simulation in the CMD using $'+'$ symbols for the relaxed
criterion and $'\times'$ symbols for the strict criterion.
We also plot a typical simulated light curve for a
high-magnification microlensing event in Fig.\ref{light}. The error
bars are $1\sigma $ results from the Poisson fluctuations in the
number of photons received by the detector, according to equation
(\ref{sss}), shifted by a Gaussian function from the theoretical
light curve. The top panel of the Fig. \ref{light} represents the
peak of the theoretical light curve and the simulated data points
with the duration of $\sim 90~min$. The whole light curve is shown
in the bottom panel. Similar to the distribution of $t_E$ towards a
given source, we plot the distribution of the relevant transit time
of $t_{1/2}$ in Fig. \ref{time}. The overall number of the events
are normalized according to Table \ref{tab2}.



One of the backgrounds that we should take into account in the
simulation of the light curves is the hot pixels of the CCD
resulting from the cosmic rays.  The probability of a hot pixel in
the {\it HST} for $1000~s$ exposure time is $1.5\sim 3 \%$
\cite{hot}. We take into account the effect of cosmic rays and
discard the hot pixels from the light curve. The other backgrounds
that can be mistaken for a microlensing events are the variable
stars and novas. The annual average Novas in M87 is about $\sim 64$
\cite{nova87}. So, during the observations over 1 or 2 d, we can
expected one event, at most, to be detected. Also, the variable
stars do not have enough variation in the light curve to be detected
in the high blending medium. We ignore the effects of novas and
variable stars.


\begin{figure}
\psfig{file=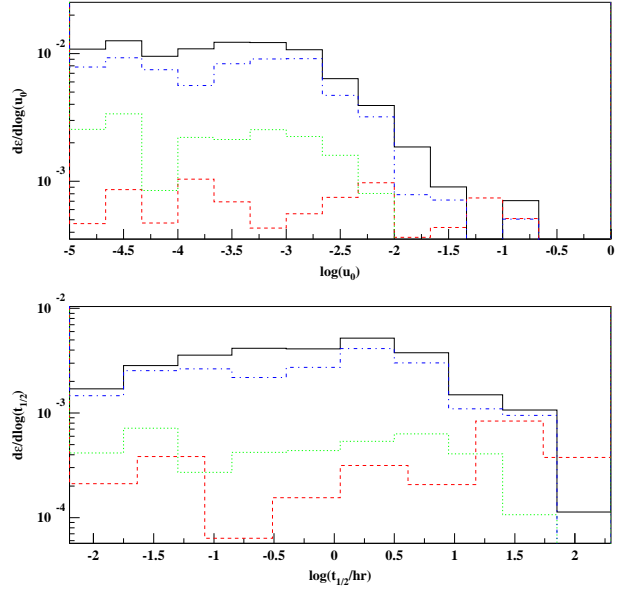,angle=0,width=9.cm,clip=} \caption{The
detection efficiency as function of the $log(u_{0})$ (top panel) and
$\log(t_{1/2}/h)$ (low panel) for two strategies of short cadence
and long cadence observations. The short cadence observation with
one day, two days and three days are shown with green dotted line,
blue dots-dashed line and black solid line, respectively. The
efficiency for the long cadence observation with the duration of one
month is shown with red dashed line. \label{det}}
\end{figure}

\begin{table*}
\begin{center}
\begin{tabular}{|c|c|c|c|c|c|c|c|}
& {$T_{obs}$ $(day)$ } & {$1 $} & {$2 $} &
{$3 $} & {$30 $}\\
\hline\hline
1 &$N_{e} (1/day)$& 4.2 & 13.4 & 18.5 & 0.02 & \\
\hline
2 &$t_{1/2}(hr)$& 16.3 & 20.9 & 22.6 & 62.4 & \\
\hline
3 &$t_{E} (day)$& 22.8 & 21.4 & 21.8 & 39.9 & \\
\end{tabular}
\end{center}
\caption{ The first three columns are for one, two and three days
observation. The forth column is the result of simulation for the
observation with the duration of $30$ days with the strategy of
taking 30 data points during one month. All the simulation is done
based on using WFC3 HST camera, except the last one with WFC2. The
first row shows the number of events per one day. The second row
indicates the average transit time of the events in hour and the
third row indicates the average Einstein crossing time in
day.\label{tab3} }
\end{table*}


\section{RESULT}
\label{results} In this section we report the results of the
Monte-Carlo simulation as the number of events and the mean value
for the parameters of the lenses. For each structure along the line
of sight, we define a threshold form the simulation where a
microlensing event is detected in the range of $u_0<u_{th}$. The
overall number of events can be obtained by summing the number of
events for each structure as follows:
\begin{equation}
\frac{N_{e}}{T_{obs}} = 4 \sum_{i=1}^{N_{s}} \int_\Omega {\tau^{i}
\left<\frac{\tilde{t_{E}^i}}{\epsilon^i(\tilde{t_{E}})}\right>^{-1}
{u^{i}_{th}(\theta)}^2 \frac{dN(\theta)}{d\theta}\theta d\theta}.
\end{equation}
Here, $N_s$ is the number of the structures, the superscript $'i'$
corresponds to the $i$th structure along the line of sight and
$\epsilon^i(\tilde{t_{E}})$ corresponds to the detection efficiency.
$<{\tilde{t_{E}^i}}/{\epsilon^i(\tilde{t_{E}})}>$ is the average
value for the ratio of the reduced Einstein crossing time to the
detection efficiency with the threshold impact parameter of $u_{th}$
for each structure. The integration is performed over the surface of
M87 using the column density of the stars from the model. We also
obtain the differential number of events $ {dN}/{d\theta}$ as a
function of angular separation from the center of the structure, as
represented by the dashed line in Fig.\ref{event}. Unlike the
results from the rough estimation, the results from the Monte Carlo
simulation show that the central part of the structure is better for
pixel-lensing. Our simulation results in $4.2$ pixel lensing for $1$
d of observations with the {\it HST}. The corresponding parameters
of the events are shown in Table \ref{tab2}. In order to test how
sensitive the detection is to the duration of the observation, we
run the Monte-Carlo simulation for the duration of $1$, $2$ and $3$
d using both with the new strategy and that of Baltz et al. (2004)
for one month. We report the daily numbers of events in Table
(\ref{tab3}).

The results for the observational efficiency is shown in Fig
\ref{det}, where we plot the detection efficiency in terms of
$log(u_{0})$ (top panel) and $log(t_{1/2})$ (bottom panel) for two
observational strategies and four observational durations. The
minimum impact parameter is limited by the finite-size effect below
$log(u_0) = -2.5$, where the efficiency remains constant. The
efficiency diagrams predict that intensive observations using the
{\it HST} over one or a few days, with the duration of the orbital
time, is much more efficient than observations over one month,
taking data point per day. The new strategy increases the
observational efficiency by almost one order of magnitude. Table
\ref{tab3} summarize the results of simulation for various
observational strategies.

\section{conclusion}
\label{conc} Gould (1995) proposed an observational strategy for
pixel lensing of the Virgo cluster using the {\it HST}, taking one
data point per day for the duration of one month. Baltz et al.
(2004) performed the observations with the WFC2 camera of the HST
and found seven variable candidates, from which one was shown to be
a microlensing candidate. They used one microlensing candidate and
put limit on the contribution of MACHOs in the Virgo halo.

In this paper, we have proposed a new observational strategy,
emphasizing the detection of the shor-duration very high
magnification events. To detect such events, we need high cadence
observation by HST, which means one observation per orbit for the
duration of one to few days. First, we made a rough estimation of
the number of observed events using this strategy. For the detailed
study, we performed a Monte-Carlo simulation to obtain the number of
pixel lensings events.
With {\it HST} observation over $1$ d, and considering $20$ per cent
of the Virgo halo to be composed of MACHOs, we expect to see $\sim
4.2$ events. If we look at the result of the observations over $2$
or $3$ d, there are more than two and three times the number of the
events than for observation taken over $1$ d. We note that, with the
new strategy, taking observations over 15 orbits in 1 d uses half of
the observational time than one month of observations with the
former observational strategy, and it increases the number of events
substantially. This new strategy could provide new possibility for
pixel-lensing observations on cosmological scales. \\

\textbf{Acknowledgment} We would like to thank the referee for
valuable comments. We thank Mansour Karami for helping to generate
the stellar population of stars and we thank Kailash Sahu for his
guidance concerning the details of the {\it HST} observations.


\end{document}